\documentclass[conference, letterpaper]{IEEEtran}
\newcommand{\subparagraph}{}
\usepackage[top=0.748in, bottom=1in, left=0.63in, right=0.63in]{geometry}
\usepackage{etoolbox}

\makeatletter
\patchcmd{\@maketitle}{\Huge}{\fontsize{18}{21pt}\selectfont}{}{}
\renewcommand{\@biblabel}[1]{[#1]\hfill}
\makeatother

\renewcommand{\sublargesize}{\fontsize{12}{14pt}}
\usepackage{titlesec}
\titlespacing{\section}{0pt}{12pt}{6pt}
\titlespacing{\subsection}{0pt}{6pt}{6pt}

\usepackage{caption}
\newcommand{\LAMCtitle}[1]{\title{#1\vspace*{-1.8ex}}}

\usepackage[T1]{fontenc}
\usepackage[american]{babel}
\usepackage{xcolor}
\usepackage{graphicx}
\usepackage{amsmath,amssymb}
\usepackage{float}
\usepackage{multirow}
\usepackage{cite}
\usepackage{subcaption}
\usepackage[protrusion=false]{microtype}
\usepackage{times}
\usepackage{textcomp}
\usepackage{bm}

\LAMCtitle{A 188-260 GHz Distributed Frequency Multiplier\\ Achieving 4~dBm Saturated Output Power in 65\,nm CMOS}

\author{

\IEEEauthorblockN{
Ahmet Yelboğa\IEEEauthorrefmark{1},
Shuhei Amakawa\IEEEauthorrefmark{2},
Korkut Kaan Tokgöz\IEEEauthorrefmark{3}
}

\IEEEauthorblockA{
\IEEEauthorrefmark{1}\IEEEauthorrefmark{3}Faculty of Engineering and Natural Sciences, Sabancı University, Türkiye\\
\IEEEauthorrefmark{2}Graduate School of Advanced Science and Engineering, Hiroshima University, Japan
}

\IEEEauthorblockA{
\IEEEauthorrefmark{1}ahmetyelboga@sabanciuniv.edu,
\IEEEauthorrefmark{2}amakawa@hiroshima-u.ac.jp,
\IEEEauthorrefmark{3}korkut.tokgoz@sabanciuniv.edu
}

}
\begin{document}

\maketitle

%
%
\begin{abstract}
Sub-terahertz (sub-THz) low-phase-noise signal sources commonly employ frequency multiplier chains to extend the operating frequency of lower-frequency oscillators. This paper presents the design and simulations of a wideband, high-output-power frequency multiplier implemented in TSMC 65\,nm CMOS technology. The proposed design is based on a distributed frequency multiplier with a quarter-wave transmission-line termination at the drain side, where the reflected and forward-traveling waves are constructively combined to enhance the output power. The proposed frequency doubler achieves a minimum conversion loss of 12\,dB at an input frequency of 115\,GHz with an input power of 12\,dBm, resulting in an output power of 0\,dBm and 12.6\,dB fundamental suppression. For the same input power, the output power remains within 3\,dB of its peak value over the 188--260\,GHz output frequency range. At 230\,GHz, the proposed design achieves a saturated output power of 4\,dBm while occupying an area of 0.149\,mm$^{\bm{2}}$.
\end{abstract}

%
%

\section{Introduction}

The sub-terahertz (sub-THz) frequency band, particularly the 300\,GHz band, provides abundant spectrum resources for high-speed wireless communications \cite{9269931,8732419}. Among the available semiconductor technologies, CMOS is an attractive solution due to its scalability, high yield, and high integration capability. However, implementing a power amplifier after the mixer at sub-THz frequencies is impractical in CMOS because of the limited gain available from transistors. Consequently, sub-THz CMOS transmitters commonly adopt a mixer-last architecture \cite{8617468,11378924}. In the absence of a power amplifier after the mixer, minimizing the mixer conversion loss becomes essential. Therefore, a high-power signal source is required to compensate for the conversion loss of the mixer. Since phase noise generally degrades as the local oscillator (LO) frequency increases, high-frequency LO generation is commonly realized by frequency multiplication from a lower-frequency, low-phase-noise source~\cite{chen2025review}.

Compared to higher-order frequency multipliers, frequency doublers are often preferred in sub-THz applications because they offer the highest output saturated power. Thus, most previously reported sub-THz frequency doubler topologies rely on biasing the transistors in the nonlinear region, typically near the threshold voltage ($V_{\mathrm{TH}}$) \cite{abdo2017100, guo2024190, wu20170,6025224}.

As significant conversion loss is unavoidable in CMOS sub-THz frequency multipliers, traveling-wave structures were employed to combine the current contributions of multiple distributed multiplier cells, thereby maximizing the output power \cite{5997295}. However, in the referenced work, the outputs of all multiplier cells were combined at a single node, which limited the flexibility of integration with subsequent circuit blocks. In contrast, this work employs a wideband sub-THz distributed frequency multiplier with traveling-wave structures at both the input and output, offering integration flexibility while maintaining broadband operation using TSMC 65\,nm CMOS technology. The proposed frequency multiplier achieves a saturated output power of 4\,dBm at 230\,GHz.

\begin{figure}[t]
	\centering
	\includegraphics[width=1\linewidth]{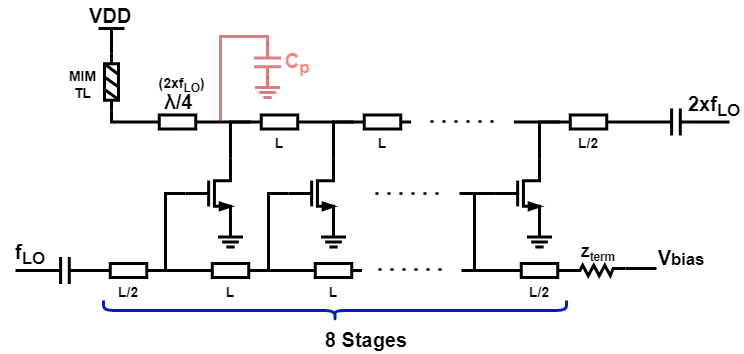}
	\caption{Distributed frequency multiplier schematic}
	\label{fig:schematic}
\end{figure}

\begin{figure}[]
	\centering
	\includegraphics[width=0.85\linewidth]{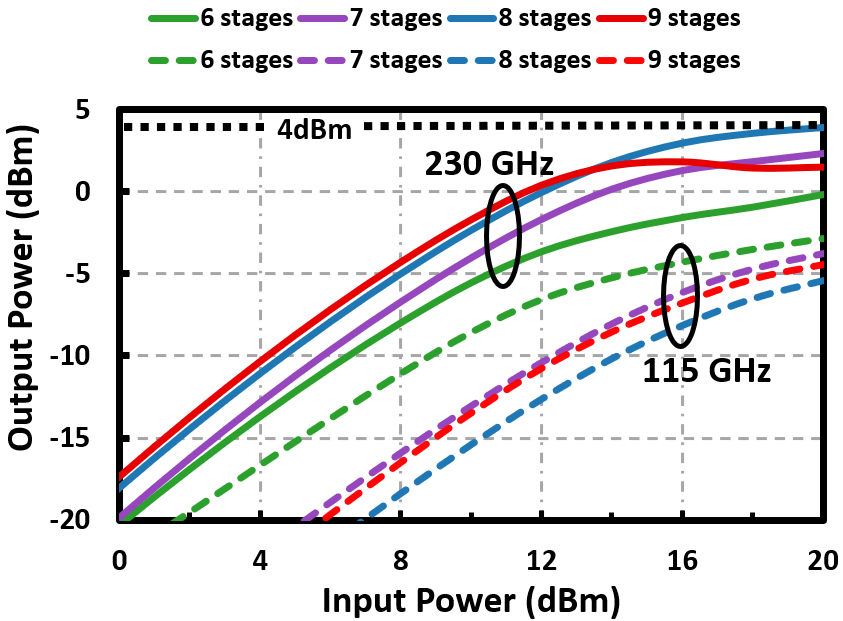}
	\caption{Second-harmonic and fundamental output power vs.\ input power at 115\,GHz for different stage counts.}
	\label{fig:stages}
\end{figure}

\section{Design Methodology}
The proposed distributed frequency multiplier, shown in Fig.~\ref{fig:schematic}, follows the distributed amplifier concept, where artificial transmission lines compensate for the gate and drain parasitic capacitances of parallel transistors to achieve ultra-wideband operation. Unlike the conventional structure, the drain side is terminated with a quarter-wave transmission line designed at 230\,GHz to obtain a reflection coefficient of $\Gamma = 1$. Consequently, the reflected wave is constructively combined with the forward-traveling wave instead of being absorbed by the termination, which boosts the output power.  A MIM transmission line \cite{amakawa2015wideband} is placed after the quarter-wave transmission line to isolate the RF signal from the DC supply. The drawback of this approach is that, unlike the conventional termination that employs an $L/2$ transmission line followed by a matched load, the quarter-wave transmission line does not cancel the parasitic capacitance of the transistor closest to the termination, resulting in poorer output matching. Since the proposed circuit is designed for the LO path, a good output matching is less critical than in data-path circuits. Therefore, the resulting mismatch does not raise inter-symbol interference concerns, and the associated degradation in matching is acceptable because the mixer performance is primarily determined by the available voltage swing at its LO input.
\begin{figure}[t]
	\centering
	\includegraphics[width=0.9\linewidth]{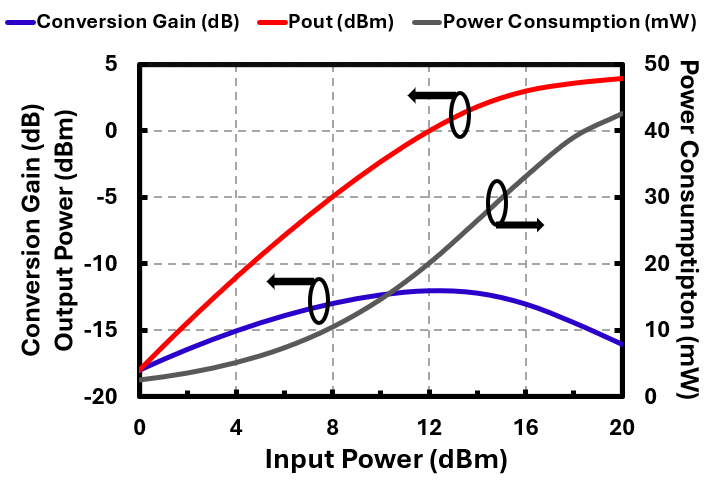}
	\caption{Conversion gain, output power, and DC power consumption vs.\ input power at 115\,GHz.}
	\label{fig:cl_pc}
\end{figure}
\begin{figure}[t]
	\centering
	\includegraphics[width=0.82\linewidth]{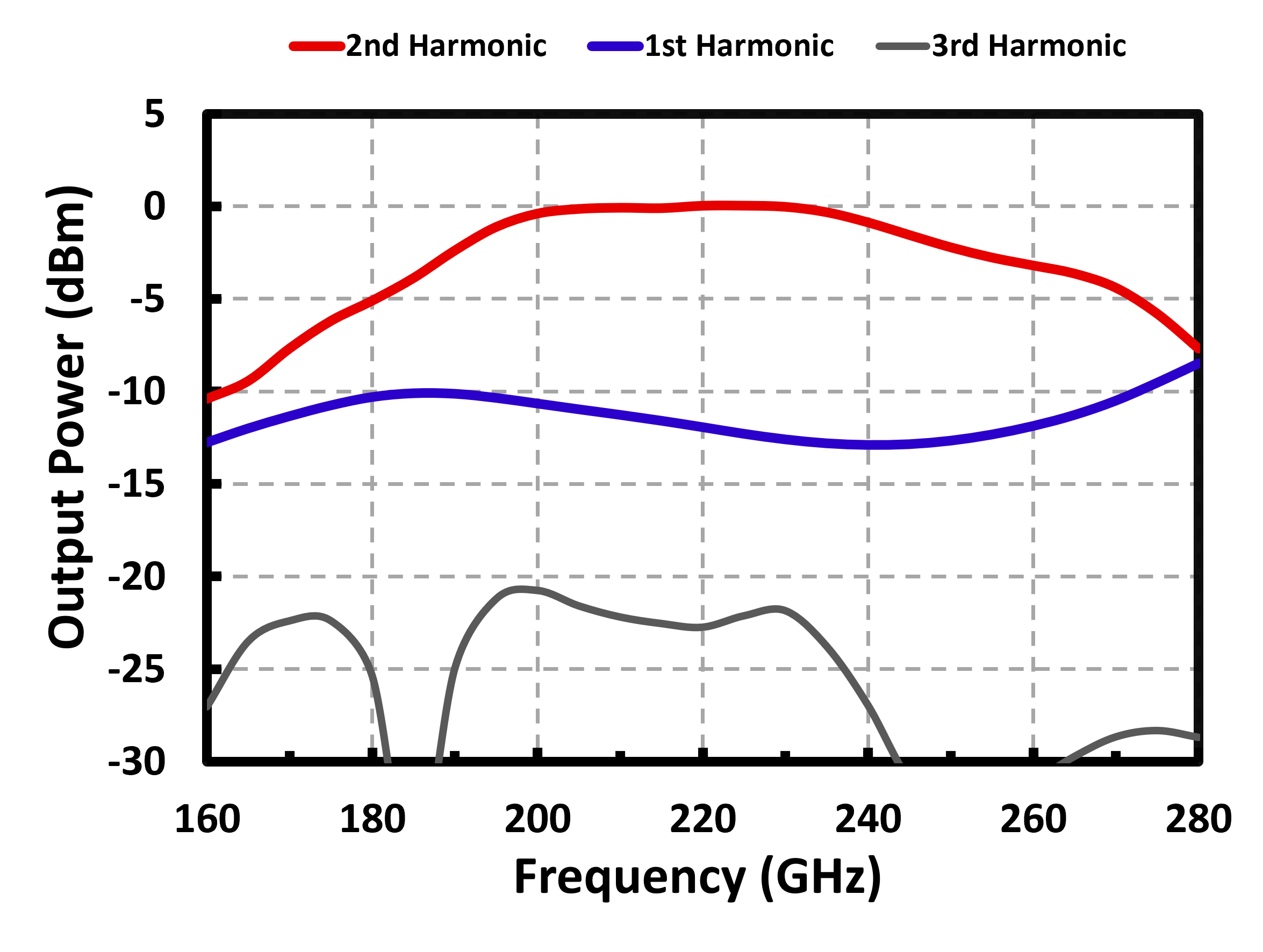}
	\caption{Harmonic output powers vs. output frequency at 12\,dBm input power.}
	\label{fig:out_freq}
\end{figure}

The frequency multiplier uses class-C biased transistors with $W/L=12$\,{\textmu}m$/0.06$\,{\textmu}m under a $1\,\mathrm{V}$ supply. The transistor sizing is determined to maximize $g_\mathrm{m2}$ while keeping $g_\mathrm{m}$ at an acceptable level, resulting in an optimum total transistor width of $12$\,{\textmu}m for maximum output power over the target frequency range. After determining the transistor dimensions, the transmission line length is optimized to $L=50$\,{\textmu}m to compensate for the transistor parasitic capacitances. The optimum number of stages is determined by sweeping the input power at $115\,\mathrm{GHz}$ and comparing the generated fundamental and second-harmonic output powers. As shown in Fig.~\ref{fig:stages}, the eight-stage design is selected because it exhibits the highest output saturated power and the highest fundamental suppression among the evaluated configurations. The input and output DC-blocking fringe capacitors are implemented in the M8 metal layer using HFSS EM simulations to ensure accurate modeling. 

\begin{figure}[t]
    \centering

    \begin{subfigure}[b]{0.43\linewidth}
        \centering
        \includegraphics[width=\linewidth]{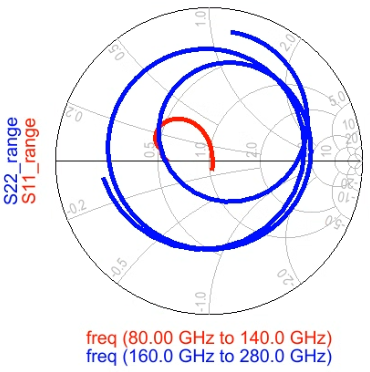}
        \caption{Smith chart}
        \label{fig:matching_smith}
    \end{subfigure}
    \hfill
    \begin{subfigure}[b]{0.55\linewidth}
        \centering
        \includegraphics[width=\linewidth]{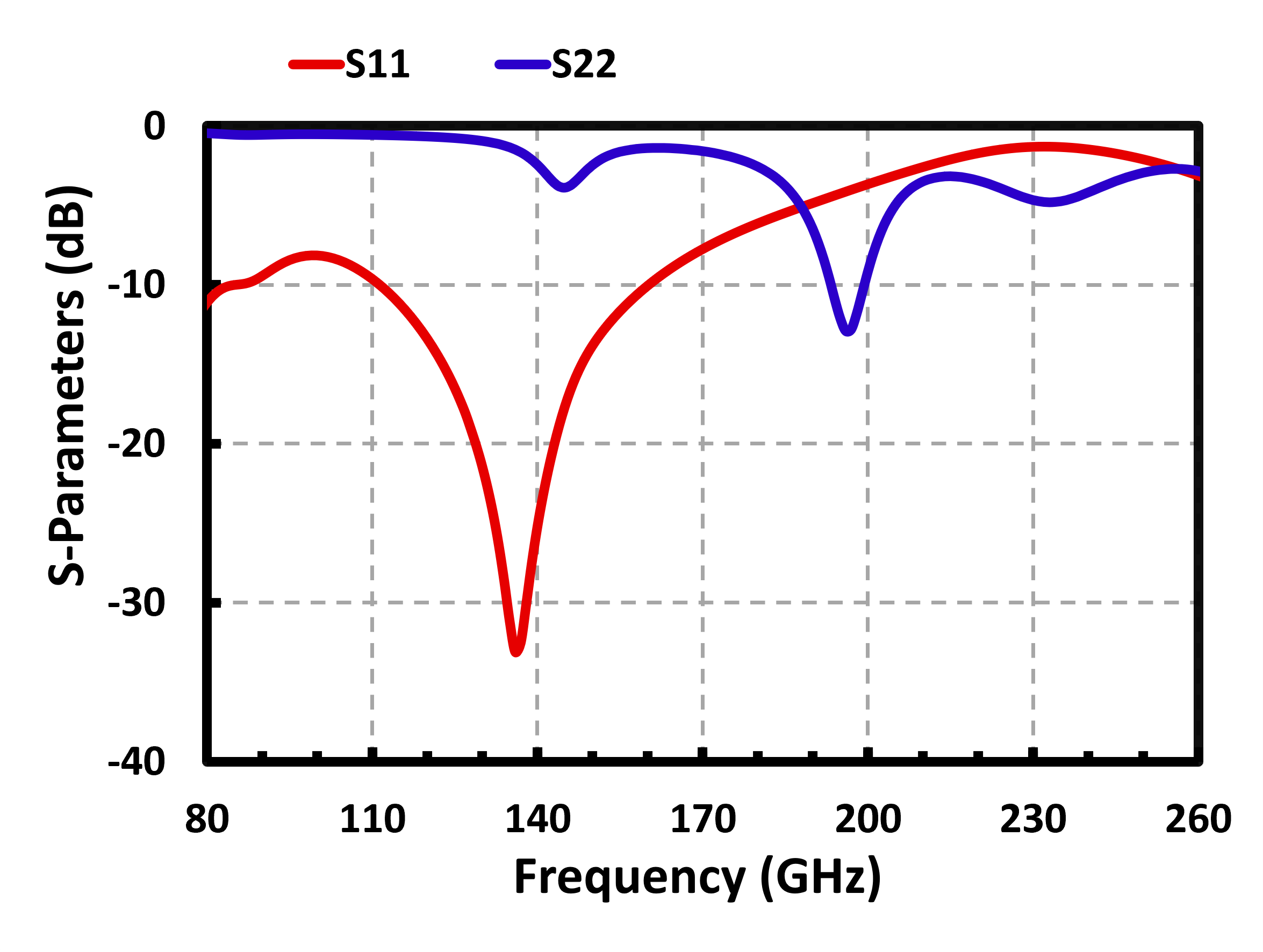}
        \caption{Rectangular plot}
        \label{fig:matching_rect}
    \end{subfigure}

    \caption{Input ($S_{11}$) and output ($S_{22}$) reflection coefficients shown as (a) Smith chart over the corresponding operating frequency ranges (80--140~GHz for $S_{11}$ and 160--280~GHz for $S_{22}$) and (b) rectangular plots over the 80--260~GHz frequency range.}
    \label{fig:matching}
\end{figure}

\section{Simulation Results}
Conversion loss is unavoidable in sub-THz frequency multipliers without an amplifier stage. The proposed design achieves a minimum conversion loss of 12\,dB at an input frequency of 115\,GHz with an input power of 12\,dBm, resulting in an output power of 0\,dBm and 12.6\,dB fundamental suppression, as shown in Fig.~\ref{fig:cl_pc} and Fig.~\ref{fig:out_freq}. At the same input power, the output power remains within 3\,dB of its peak value over the 188--260\,GHz output frequency range that demonstrates  a wideband operation with 32\% fractional bandwidth (Fig.~\ref{fig:out_freq}). Within the operating band, the fundamental suppression remains greater than 10\,dB, while the third-harmonic suppression remains greater than 20\,dB. At 230\,GHz, the design achieves a saturated output power ($P_{\mathrm{sat}}$) of 4\,dBm with an input power of approximately 20\,dBm. As the input power increases, the conduction angle of the class-C biased transistors increases, which leads to higher DC power consumption and a reduction in DC efficiency, as shown in Fig.~\ref{fig:cl_pc}. The input and output return losses are presented in Fig.~\ref{fig:matching}. The relatively poor output return loss results from the quarter-wave termination, as the parasitic capacitance of the transistor closest to the termination is intentionally left uncompensated to boost the output power. All results in this work were obtained through simulations using Ansys HFSS and Keysight ADS. The material properties used in the HFSS simulations were obtained from the PDK.

\renewcommand{\thetable}{\Roman{table}}
\begin{table*}[!t]
\caption{\textsc{PERFORMANCE COMPARISON}}
\label{tab:sota}
\small
\centering
\sf

\begin{tabular}{|c|c|c|c|c|c|c|c|}
\hline
\textbf{Ref.} &
\textbf{Type} &
\begin{tabular}[c]{@{}c@{}}\textbf{Output}\\\textbf{Frequency}\\\textbf{(GHz)}\end{tabular} &
\begin{tabular}[c]{@{}c@{}}\textbf{Conversion Gain}\\\textbf{(dB)}\end{tabular} &
\begin{tabular}[c]{@{}c@{}}\textbf{$\mathbf{P}_{\mathrm{sat}}$}\\\textbf{(dBm)}\end{tabular} &
\begin{tabular}[c]{@{}c@{}}\textbf{$\mathbf{P}_{\mathrm{DC}}$}\\\textbf{@$\mathbf{P}_{\mathrm{sat}}$}\\\textbf{(mW)}\end{tabular} &
\begin{tabular}[c]{@{}c@{}}\textbf{Area}\\\textbf{($\,\mathbf{mm}^2$)}\end{tabular} &
\textbf{Technology}
\\
\hline

\multicolumn{7}{|c|}{\textbf{Frequency Multipliers}}\\
\hline

This work* & Doubler & 188--260 & -12 & 4 & 42.5 & 0.149 & 65\,nm CMOS \\
\hline

\cite{guo2024190} & Doubler & 190--217 & -7.5 & 1.3& 33** & - & 65\,nm CMOS\\
\hline

\cite{6025224} & Doubler & 170--200 & -6.4 & 0 & 33***& 0.34 & 45\,nm SOI CMOS\\
\hline

\cite{5997295} & Doubler & 220--275 & -11.4 & 0 & 40 & 0.022 & 65\,nm CMOS\\
\hline

\multicolumn{7}{|c|}{\textbf{Frequency Multiplier Chains}}\\
\hline

\cite{abdo2017100} & Doubler + Harm. Rej. + Amp.  & 100-123 & 17.2 & 5.5 & 116 & 0.832 & 65\,nm CMOS\\
\hline

\cite{guo2024190} & DA + Doubler & 190-217 & 13.2 & 0.3 & 100 & 0.49& 65\,nm CMOS \\
\hline

\cite{wu20170} & Inj. Lock. Amp + Doubler & 185-202 & 1.7**  & 1.92 & 96 & 0.176 & 65\,nm CMOS\\
\hline

\cite{9796560} & DA + Doubler & 208--233 & -7 & -4.9 & 26.4 & 0.2 & 28\,nm CMOS\\
\hline

\cite{sharma2016160} & Amp + DA + Doubler & 160-310 & 3 & 3 & 80 & 0.713 & 65\,nm CMOS\\
\hline

\multicolumn{7}{l}{$^{*}$ Simulated results.  $^{**}$ Extracted from figure.    $^{***}$ Calculated result.}\\

\end{tabular}
\end{table*}

\begin{figure}[]
	\centering
	\includegraphics[width=1\linewidth]{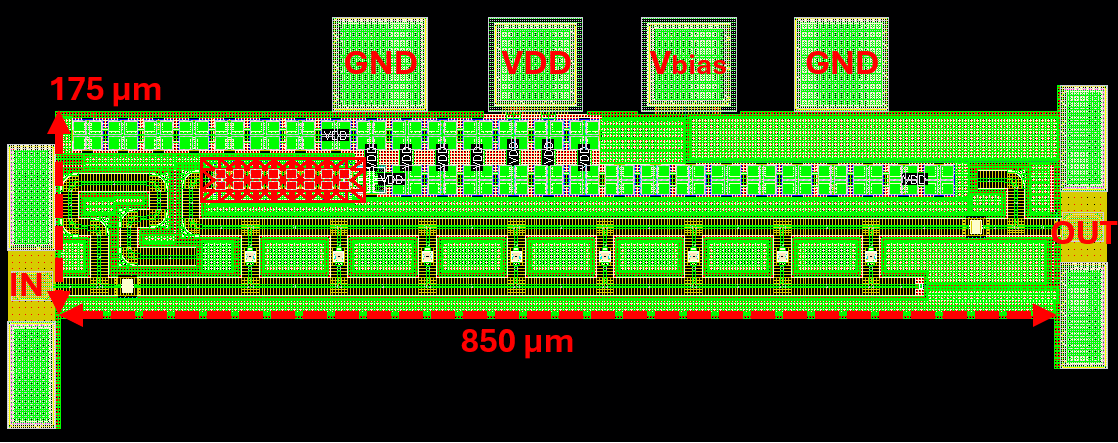}
	\caption{Layout of the designed distributed frequency multiplier}
	\label{fig:layout}
\end{figure}

The designed layout, which is shown in Fig.~\ref{fig:layout}, occupies an area of $0.149\,\mathrm{mm}^2$ with dimensions of $175\,\mu\mathrm{m} \times 850\,\mu\mathrm{m}$, excluding the pad area.

Table~\ref{tab:sota} compares the simulation results of the proposed design with previously reported standalone frequency multipliers and frequency multiplier chains. Since relatively few standalone frequency multipliers operating in the sub-THz range have been reported, multiplier chains are also included to provide a broader perspective. Among the standalone frequency multipliers, the proposed design achieves the highest saturated output power while maintaining a competitive conversion loss. Furthermore, it provides one of the widest 3\,dB output power bandwidths.

\section{Conclusion}
This paper presented a distributed frequency doubler for sub-terahertz signal generation in TSMC 65\,nm CMOS technology. The proposed architecture employs a quarter-wave transmission-line termination at the drain side to constructively combine the reflected and forward-traveling waves, enhancing the output power while facilitating integration with subsequent circuit blocks. Simulation results demonstrate a minimum conversion loss of 12\,dB at 115\,GHz, a 3\,dB output power bandwidth from 188--260\,GHz, and a saturated output power of 4\,dBm at 230\,GHz. These results confirm that the proposed architecture can provide sufficient output power over a wide bandwidth to be used as a source for sub-terahertz mixers.

\vspace{8mm}
\bibliographystyle{IEEEtran}

\bibliography{IEEEabrv,IEEEexample}

\end{document}